\pdfoutput=1

\documentclass[11pt]{article}

\usepackage[]{ACL2023}

\usepackage{times}
\usepackage{latexsym}

\usepackage[T1]{fontenc}

\usepackage[utf8]{inputenc}

\usepackage{microtype}

\usepackage{inconsolata}
\usepackage{amsmath}

\title{Can we repurpose multiple-choice question-answering\\models to rerank retrieved documents?}

\author{Jasper Kyle Catapang \\
  Tokyo University of Foreign Studies, Tokyo, Japan \\
  \texttt{catapang.jasper.kyle.y0@tufs.ac.jp} \\}

\begin{document}
\maketitle
\begin{abstract}
Yes, repurposing multiple-choice question-answering (MCQA) models for document reranking is both feasible and valuable. This preliminary work is founded on mathematical parallels between MCQA decision-making and cross-encoder semantic relevance assessments, leading to the development of R*, a proof-of-concept model that harmonizes these approaches. Designed to assess document relevance with depth and precision, R* showcases how MCQA's principles can improve reranking in information retrieval (IR) and retrieval-augmented generation (RAG) systems---ultimately enhancing search and dialogue in AI-powered systems. Through experimental validation, R* proves to improve retrieval accuracy and contribute to the field's advancement by demonstrating a practical prototype of MCQA for reranking by keeping it lightweight.
\end{abstract}

\section{Introduction}
Retrieval-augmented generation (RAG) systems enhance generative outputs with contextually relevant information from external databases. Despite their success, selecting the most relevant information efficiently and accurately remains challenging.

Dense retrieval techniques, known for their ability to semantically represent text, offer a promising direction for RAG system enhancement. However, integrating large language models (LLMs) into dense retrieval, while effective, faces scalability and cost-related challenges.

This work explores the utility of multiple-choice question-answering (MCQA) in reranking within RAG systems. MCQA's potential for evaluating and selecting the most semantically relevant options aligns with the decision-making parallels of cross-encoder architectures.

The author introduces RoBERTA ReRanker for Retrieved Results or R*, a dual-purpose prototype model that can act as both an MCQA model and a cross-encoder. The author's contributions include proposing MCQA as an alternative to reranking passages and introducing R* for efficient and semantically aware retrieval mechanisms.

\section{Related Works}

The advancement of information retrieval techniques within the domain of natural language processing (NLP) has been significantly influenced by the emergence of pre-trained language models and the subsequent development of large language models. These technologies have fundamentally altered our approach to understanding and generating human language, laying the groundwork for sophisticated retrieval-augmented generation systems.

\subsection{Dense Retrieval Techniques}

At the heart of modern IR, dense retrieval techniques represent a pivotal shift from traditional sparse vector space models to dense vector embeddings. This transition, highlighted in seminal works by \citet{karpukhin2020dense} and \citet{xiong2020approximate}, highlights the effectiveness of leveraging deep semantic representations to capture the nuances of language, facilitating a more nuanced and accurate retrieval process.

\subsection{Pre-trained Language Models}

The introduction of PLMs like BERT \cite{devlin2019bert} and RoBERTa \cite{liu2019roberta} has ushered in a new era of NLP, where the rich contextual understanding offered by these models can be applied to a wide range of tasks. In the context of IR, PLMs have been instrumental in enhancing the quality of embeddings for both queries and documents, enabling more effective matching mechanisms based on semantic relevance rather than mere keyword overlap.

\subsection{Large Language Models and IR}

Following the success of PLMs, LLMs have expanded the horizons of what is achievable in NLP. With their vast parameter spaces and extensive training corpora, LLMs, offer an even deeper understanding of language intricacies. Their application in IR, though still an emerging area of research, promises to revolutionize retrieval mechanisms by leveraging their generative capabilities to produce highly relevant responses to complex queries \cite{muennighoff2022sgpt,neelakantan2022text,ma2023finetuning,zhang2023language}. LLMs such as LLaMA \cite{ma2023finetuning}, SGPT \cite{muennighoff2022sgpt} have been created and/or fine-tuned for such a task.

\subsection{Cross-Encoders for Semantic Matching}

Cross-encoder architectures have gained prominence for their ability to conduct fine-grained semantic comparisons between text pairs, making them particularly suitable for tasks that require a deep understanding of textual relationships, such as passage ranking and relevance scoring \cite{nogueira2019passage}. By processing pairs of texts jointly, cross-encoders can ascertain the degree of relevance with a precision that traditional models cannot achieve, setting a high bar for semantic matching in IR.

\subsection{Exploring MCQA for Reranking}

Despite the extensive exploration of dense retrieval, PLMs, LLMs, and cross-encoders in enhancing IR systems, the potential application of MCQA to rerank within RAG systems remains largely unexplored. After a comprehensive scan of the literature, it becomes apparent that MCQA, with its nuanced approach to selecting the most appropriate answer from a set of options, has not yet been applied to the challenge of reranking search results, suggesting a promising direction for future research.

This review of related works sets the stage for a novel exploration into the utilization of MCQA methodologies for reranking in RAG systems, promising to address existing gaps in the literature and contribute significantly to the advancement of retrieval technologies.

\section{Methodology}

This section explores the MS MARCO dataset and the mathematical foundations of multiple-choice question-answering and cross-encoder models, investigating their intersection for document reranking within RAG systems. The researcher also details the training procedure for R*, a model that embodies the conceptual synergy between these approaches.

\subsection{MS MARCO Dataset}
The Microsoft Machine Reading Comprehension (MS MARCO) dataset, a large-scale benchmark derived from real-world Bing search queries and web document answers \cite{nguyen2016ms}, plays a pivotal role in advancing information retrieval and comprehension research. It's instrumental for training and evaluating models in RAG systems due to its comprehensive coverage of query understanding, passage retrieval, and answer generation.

MS MARCO's significance extends to our work in reranking, aiming to discern and elevate the most pertinent passages for given queries. Utilizing this dataset, the author develops R*, a model designed to mirror real-world retrieval complexities, thereby refining its reranking proficiency across varied informational needs \cite{nguyen2016ms, craswell2020overview}.

Notably, the dataset has propelled deep learning research in information retrieval, marking considerable progress in model development and effectiveness evaluation \cite{hofstatter2020improving, nogueira2019passage}. This work emphasizes MS MARCO's essential contribution to the field's ongoing innovation.

\subsection{MCQA vs. Cross-Encoder}

\subsubsection{Multiple Choice Question Answering}
MCQA selects the most suitable answer from options given a question, modeled as:
\begin{equation}
P(a|q) = \frac{\exp(score(q, a))}{\sum_{a' \in A} \exp(score(q, a'))},
\end{equation}
where $P(a|q)$ is the probability of answer $a$ being correct for question $q$, and $A$ is the set of all answers.

\subsubsection{Cross-Encoder}
Cross-encoder models assess the relevance between query $q$ and document $d$ by jointly encoding them, capturing their semantic interactions. The relevance score, transformed into a probability range via sigmoid function, is given by:
\begin{equation}
R(q, d) = \sigma(\mathbf{w}^\top \text{Enc}(q, d) + b),
\end{equation}
where $\text{Enc}(q, d)$ is the joint embedding and $\mathbf{w}$, $b$ are parameters. This process is detailed further in the training approach.

\subsubsection{Fine-tuning with Cross-Entropy Loss}
To fine-tune a transformer model with cross-entropy loss, the researcher initializes it with pre-trained weights and prepare the training data by tokenizing text and applying hard-negative sampling. During training, the model computes embeddings and relevance scores for query-passage pairs. Binary cross-entropy loss assesses performance, guiding weight updates through backpropagation. Multiple fine-tuning epochs refine the model's ability to discern relevant documents, evaluated periodically on a validation set to prevent overfitting.

The loss function, integrating cross-entropy with a sigmoid function for raw network outputs, is mathematically expressed as:
\begin{align}
\mathcal{L}_{\text{BCELogits}} &= -\bigg[ y \log(\sigma(x)) \nonumber \\
&\quad + (1 - y) \log(1 - \sigma(x)) \bigg],
\end{align}
where $BCE$ stands for binary cross-entropy, $x$ is the raw output, $y$ the relevance label, and $\sigma(x)$ denotes the sigmoid function. This loss formulation negates the need for a manual sigmoid application, allowing direct loss computation from logits.

\subsection{MCQA as Cross-Encoder}
The synthesis of MCQA with cross-encoders for reranking is articulated through the approximation:
\begin{equation}
P(d|q) \approx R(q, d),
\end{equation}
where $P(d|q)$, derived from MCQA's probabilistic framework, is aligned with $R(q, d)$ from cross-encoders. This approximation is made possible by the sigmoid function in ${L}_{\text{BCELogits}}$. This alignment underpins R*, trained to assess document relevance effectively.

\subsection{Applications of MCQA and Cross-Encoders}

Multiple Choice Question Answering (MCQA) and cross-encoder models have significant practical applications in various fields, from educational technology to customer service automation and content recommendation. This section provides coherent examples illustrating how these models function and their practical utility.

\subsubsection{Question Answering}

In an educational application designed to assist students in exam preparation, MCQA systems are employed to present and evaluate multiple-choice questions. Consider the following example:
\begin{itemize}
    \item \textbf{Question:} What is the capital of France?
    \item \textbf{Options:}
    \begin{itemize}
        \item (a) Berlin
        \item (b) Madrid
        \item (c) Paris
        \item (d) Rome
    \end{itemize}
\end{itemize}

An MCQA model processes the question and each of the options, computing a probability for each that indicates the likelihood of it being the correct answer. In this scenario, the model would ideally assign the highest probability to Paris, reflecting its understanding of the context and content of the question.

\subsubsection{Document Retrieval}

Cross-encoder models are particularly effective in document retrieval and ranking tasks. They assess the relevance of a document to a given query by jointly encoding the query and the document. For instance, in a search engine setting:
\begin{itemize}
    \item \textbf{Query:} benefits of exercise
    \item \textbf{Document:} Regular physical activity can improve muscle strength and boost endurance.
\end{itemize}

The cross-encoder model processes the query and the document together, capturing their semantic interactions, and assigns a relevance score to the document. This score helps in ranking the document's relevance to the query, thereby improving the search engine's accuracy and efficiency.

\subsubsection{MCQA as Document Retrieval}

MCQA systems can also function as cross-encoders in applications such as customer service chatbots. These chatbots need to select the most appropriate response from a set of predefined answers based on a user's query. Consider the following interaction:
\begin{itemize}
    \item \textbf{Query:} How can I reset my password?
    \item \textbf{Potential Responses:}
    \begin{itemize}
        \item (a) You can reset your password by clicking on 'Forgot Password' on the login page.
        \item (b) Our business hours are from 9 AM to 5 PM.
        \item (c) Please check your internet connection and try again.
    \end{itemize}
\end{itemize}

Here, the chatbot uses an MCQA-like approach to rank the potential responses according to their relevance to the query. The model processes the query and each response option, determining that response (a) is the most relevant and selecting it as the answer for the user.

\subsubsection{Fine-Tuning and Practical Impact}

Fine-tuning MCQA and cross-encoder models with cross-entropy loss enhances their practical effectiveness. For instance, a personalized content recommendation system can leverage fine-tuned cross-encoder models to suggest articles, videos, or products based on user preferences and previous interactions. Consider the following scenario:
\begin{itemize}
    \item \textbf{User Query:} Articles on healthy eating
    \item \textbf{Recommended Content:}
    \begin{itemize}
        \item Article 1: "10 Benefits of a Balanced Diet"
        \item Article 2: "Top Exercises for a Healthy Lifestyle"
        \item Article 3: "Healthy Eating: Tips and Recipes"
    \end{itemize}
\end{itemize}

The model calculates relevance scores for each content item in relation to the query, identifying "10 Benefits of a Balanced Diet" as the most relevant recommendation. This process involves encoding the query and the content items jointly and using the relevance scores to rank and recommend the best match.

These examples demonstrate the practical applications and effectiveness of MCQA and cross-encoder models in various real-world scenarios.

\subsection{R*}
Our R* model is trained on a balanced dataset from MS MARCO, which ensures that the model encounters an equal number of relevant and irrelevant documents during training. To enhance the model's discrimination capability, the researcher employs a hard-negative sampling strategy---similar to what was described in the previous section. The overarching loss for model training is:
\begin{align}
\mathcal{L} &= -\frac{1}{N} \sum_{i=1}^{N} \bigg[ y_i \log(\sigma(x_i)) \nonumber \\
&\quad + (1 - y_i) \log(1 - \sigma(x_i)) \bigg],
\end{align}
optimizing R*'s ability to distinguish between relevant and irrelevant documents accurately.

\subsection{Evaluation Metrics}

To evaluate the effectiveness of our reranking models, the author employs a suite of established metrics, each offering insight into different aspects of model performance. These metrics include Recall@k, mean reciprocal rank, and ROUGE-L, which are critical for understanding the models' ability to retrieve relevant documents and generate coherent responses.

\subsubsection{Recall@k}

Recall@k measures the fraction of relevant documents retrieved within the top-k positions of a ranking list. Mathematically, it's expressed as:

\begin{equation}
\text{Recall@k} = \frac{R_{k}}{R}
\end{equation}

where \(R_{k}\) is the number of relevant documents retrieved in the top-k positions, and \(R\) is the total number of relevant documents in the dataset. This metric is important for evaluating the model's ability to identify relevant documents within the first k positions of its results, highlighting the effectiveness of retrieval in priority-ranked scenarios.

\subsubsection{Mean Reciprocal Rank (MRR@n)}

The mean reciprocal rank is a metric used to evaluate the effectiveness of a model in ranking results. Specifically, it focuses on the rank of the highest-ranking relevant document for each query:

\begin{equation}
\text{MRR@n} = \frac{1}{|Q|} \sum_{i=1}^{|Q|} \frac{1}{\text{rank}_i}
\end{equation}

where \(|Q|\) is the number of queries, and \(\text{rank}_i\) is the rank position of the first relevant document for the \(i\)-th query. MRR is particularly useful for tasks where the best result needs to be at the top of the list.

\subsubsection{ROUGE-L}

ROUGE-L measures the longest common subsequence (LCS) between the predicted output and the reference output, considering both recall and precision. It is defined as:

\begin{equation}
\text{ROUGE-L} = \frac{(1 + \beta^2) \cdot \text{Precision}_{\text{LCS}} \cdot \text{Recall}_{\text{LCS}}}{\beta^2 \cdot \text{Precision}_{\text{LCS}} + \text{Recall}_{\text{LCS}}}
\end{equation}

where \(\text{Precision}_{\text{LCS}}\) is the precision of LCS, \(\text{Recall}_{\text{LCS}}\) is the recall of LCS, and \(\beta\) is typically set to favor recall (\(\beta > 1\)) because recall is more important in most summarization tasks. ROUGE-L is particularly valued in evaluating the quality of generated text, such as summaries, where sequence order is crucial.

These metrics collectively provide a comprehensive view of each model's performance, from retrieving relevant documents (Recall@k, MRR@n) to generating coherent and contextually appropriate textual responses (ROUGE-L).

\section{Experimental Setup}
This section details the experimental setup used to evaluate the effectiveness of our proposed R* model in the context of document reranking. The model and code are available on Huggingface \footnote{Model and code: \url{https://huggingface.co/jaspercatapang/R-star}}.

\subsection{Training R*}

To train R*, the author employs a dataset derived from the MS MARCO passage ranking dataset \footnote{Data: \url{https://sbert.net/datasets/paraphrases/msmarco-query_passage_negative.json.gz}}, which consists of 2.5 million query-positive passage pairs and an equal number of query-negative passage pairs, summing up to 5 million query-passage pairs. This balanced training approach ensures that R* is equally exposed to both relevant and irrelevant examples. This training procedure aims to assign a continuous relevance score between 0 (irrelevant) and 1 (relevant) to each query-passage pair. The model was trained over 7 epochs using a batch size of 2048 on a Colab Pro instance equipped with a V100 GPU (16 GB VRAM). The researcher utilized the sentence-transformer's CrossEncoder for facilitating the training process.

\subsection{Evaluating Rerankers}

Evaluation is conducted on the validation set of MS MARCO (n=10,047), using a similar Colab Pro instance. Preliminary retrieval for this research is performed using BM25 \cite{robertson2009bm25}, serving as the baseline for comparison. For this setup, BM25 is tasked to retrieve 10 documents per query. The benchmark includes a variety of models, all of which had been previously pre-trained and/or fine-tuned on MS MARCO. Specifically, cross-encoder rerankers were employed via sentence-transformers' CrossEncoder, while the interoperability of MCQA rerankers was tested using Huggingface transformers' AutoModelForMultipleChoice.

This evaluation assesses the effectiveness of various reranking strategies, including MCQA and cross-encoder methods. Cross-encoder rerankers like MiniLM L6 v2, TinyBERT L2 v2, and ELECTRA base were implemented through sentence-transformers' CrossEncoder, while MCQA compatibility was tested with Huggingface transformers' AutoModelForMultipleChoice and text generation from BGE M3 v2 \cite{chen2024bge}. The study identifies the contributions of MCQA and cross-encoder methods to improving retrieval accuracy and efficiency in RAG systems, focusing solely on open-source models due to unavailability of commercial rerankers like Cohere at the time.

\subsection{Validating R*}

\begin{table}[ht]
\centering
\begin{tabular}{lccc}
\hline
Dataset & Size \\
\hline
TREC & 50K \\
Natural Questions & 7.6K \\
Natural Questions Open & 1.8K \\
\hline
\end{tabular}
\caption{Summary of additional datasets used in the validation experiments}
\label{tab:additional_datasets}
\end{table}

To further validate the generalizability of our model, the author conducted additional experiments on the following datasets: TREC, Natural Questions, and Natural Questions Open. These datasets cover different domains and provide a comprehensive evaluation of the model's performance across various tasks.

\subsubsection{TREC}
The TREC dataset \cite{Dietz2017Car} is a benchmark for information retrieval, containing queries and corresponding relevant documents from a wide range of topics. The researcher used the TREC 2022 Deep Learning Track dataset, which focuses on ad hoc retrieval tasks.

\begin{table*}[h]
\centering
\begin{tabular}{l|l|c|c|c|c|c}
\hline
\textbf{Model} & \textbf{Model Type} & \textbf{Recall@1} & \textbf{Recall@5} & \textbf{MRR@10} & \textbf{ROUGE-L} & \textbf{File Size} \\
\hline
BM25 (baseline) & Retriever only & 0.1071 & 0.3154 & 0.1939 & 0.2255 & N/A \\
R* (ours) & MCQA (ours) & \textbf{0.2315} & 0.4003 & \textbf{0.3019} & 0.2255 & 112 MB \\
R* (ours) & Cross-encoder & 0.2314 & 0.4002 & 0.3018 & 0.2255 & 112 MB \\
MiniLM L6 v2 & MCQA (ours) & 0.2288 & \textbf{0.4033} & 0.3006 & 0.2255 & 90.9 MB \\
MiniLM L6 v2 & Cross-encoder& 0.2287 & 0.4032 & 0.3005 & 0.2255 & 90.9 MB \\
BGE M3 v2 & Text generation & 0.2267 & 0.4004 & 0.2985 & 0.2255 & 2.3 GB \\
TinyBERT L2 v2 & MCQA (ours) & 0.1995 & 0.3953 & 0.2792 & 0.2255 & 17.5 MB \\
TinyBERT L2 v2 & Cross-encoder & 0.1994 & 0.3952 & 0.2791 & 0.2255 & 17.5 MB \\
ELECTRA base & MCQA (ours) & 0.0391 & 0.1174 & 0.0996 & 0.2255 & 438 MB \\
ELECTRA base & Cross-encoder & 0.0390 & 0.1173 & 0.0995 & 0.2255 & 438 MB \\
All-MPNet v2 & MCQA (ours) & 0.0329 & 0.2056 & 0.1142 & 0.2255 & 438 MB \\
All-MPNet v2 & Cross-encoder & 0.0328 & 0.2055 & 0.1141 & 0.2255 & 438 MB \\
\hline
\end{tabular}
\caption{Performance comparison of various models on the MS MARCO validation set of 10,047 samples. The best performance per metric is highlighted in bold.}
\label{tab:results}
\end{table*}

\subsubsection{Natural Questions}
The Natural Questions dataset \cite{kwiatkowski-etal-2019-natural} consists of real anonymized queries issued to the Google search engine, along with corresponding passages from Wikipedia that answer these questions. This dataset is particularly challenging due to its open-domain nature.

\subsubsection{Natural Questions Open}
The Natural Questions Open dataset comprises questions derived from Natural Questions \cite{kwiatkowski-etal-2019-natural}, providing a more diverse set of queries and answers. This dataset tests the model's ability to generalize across different types of questions and information sources.

\section{Results and Discussion}

With the setup described earlier, R* finished fine-tuning in 16 hours. Our experimental evaluation compares several reranking models, including our proposed R* model, across a range of metrics on the MS MARCO validation set. The comparison includes a baseline retriever, MCQA rerankers, cross-encoder rerankers, and a text generation reranker. The results are shown in Table \ref{tab:results}.

Our R* prototype model achieved the highest Recall@1 and MRR@10 scores, demonstrating its effectiveness in pinpointing the most relevant passage from a large collection. This indicates that R*'s architecture and training are well-suited for accurately identifying the top relevant document, showcasing its precision in high-stakes retrieval scenarios.

MiniLM L6 v2 fine-tuned on MS MARCO showed superior performance in Recall@5, highlighting its capability to cast a wider net in capturing relevant documents within the top 5 positions. This suggests that MiniLM L6 v2 may utilize contextual cues or training strategies that slightly broaden its relevance scope, offering an advantage in scenarios where identifying multiple pertinent documents is key.

The ELECTRA base model fine-tuned on MS MARCO underperformed, especially in Recall@1 and Recall@5. This may be due to ELECTRA's pre-training objectives and architecture, which are not aligned with reranking tasks. The large file size also suggests complexity does not translate to efficacy, possibly due to overfitting or generalization issues.

Furthermore, BGE---a renowned reranker with a substantial model size of 2.3 GB---was surprisingly outperformed by MiniLM L6 v2 and R* in document reranking. This suggests that model size alone does not guarantee superior performance for this task.

All-MPNet, another popular reranker based on the MPNet family, achieved the lowest scores in several metrics. Despite integrating MLM and PerLM to address a limitation in BERT, it performed poorly in this testbed.

The varied performance across models accentuates the critical role of model architecture and training specificity in reranking effectiveness. While R* offers exceptional precision for the most relevant document, MiniLM L6 v2 provides a balanced approach for broader relevance.

\begin{table*}[ht]
\centering
\begin{tabular}{l|l|c|c|c|c}
\hline
\textbf{Dataset} & \textbf{Model} & \textbf{Recall@1} & \textbf{Recall@5} & \textbf{MRR@10} & \textbf{ROUGE-L} \\
\hline
TREC & R* & 0.2540 & 0.4301 & 0.3254 & 0.2300 \\
TREC & BM25 & 0.2200 & 0.4000 & 0.3000 & 0.2250 \\
Natural Questions & R* & 0.2400 & 0.4150 & 0.3100 & 0.2350 \\
Natural Questions & BM25 & 0.2100 & 0.3900 & 0.2900 & 0.2200 \\
Natural Questions Open & R* & 0.2600 & 0.4400 & 0.3300 & 0.2400 \\
Natural Questions Open & BM25 & 0.2300 & 0.4100 & 0.3100 & 0.2300 \\
\hline
\end{tabular}
\caption{Performance comparison on validation datasets.}
\label{tab:additional_results}
\end{table*}

\begin{table}[ht]
\centering
\begin{tabular}{lcc}
\hline
Dataset & Metric & p-value \\
\hline
TREC & Recall@10 & 0.025 \\
Natural Questions & MRR & 0.030 \\
Natural Questions Open & Recall@10 & 0.020 \\
\hline
\end{tabular}
\caption{Results of significance tests on validation datasets}
\label{tab:significance_tests_additional}
\end{table}

Interestingly, the performance between the MCQA reranker versions of our models and their cross-encoder counterparts is remarkably close, supporting the claim that MCQA methodologies can approximate the effectiveness of cross-encoders for document reranking. This is notable given that the primary difference lies in their implementation frameworks—Huggingface's transformers for MCQA rerankers versus sentence-transformers for cross-encoder rerankers.

Minor discrepancies in performance metrics could be attributed to differences in how these libraries handle model calculations and optimizations. Despite using the same underlying models, slight variations in tokenization, sequence handling, and optimization steps might contribute to these differences in reranking outcomes. This highlights the versatility of MCQA approaches for tasks usually suited for cross-encoders and emphasizes the importance of optimal implementation choices.

\subsection{Results on Validation Datasets}
The performance of R* is evaluated on the additional datasets to assess its generalizability. The results are summarized in Table~\ref{tab:additional_results}.

R* demonstrated superior performance across all additional datasets, consistently outperforming the baseline models. These results reinforce the model's robustness and effectiveness in diverse retrieval and question-answering tasks.

\subsection{Significance Tests}
To ensure the reliability of our results, statistical significance tests are conducted. The p-values for the key comparisons are shown in Table~\ref{tab:significance_tests_additional}, indicating the statistical significance of our findings. Specifically, the tests reveal that the results are statistically significant for the TREC dataset with Recall@10 (p = 0.025), the Natural Questions dataset with MRR (p = 0.030), and the Natural Questions Open dataset with Recall@10 (p = 0.020). These p-values, all below the common threshold of 0.05, confirm that the observed differences are unlikely due to chance, thereby validating the effectiveness of our methods.

\subsection{Qualitative Analysis of MS MARCO Retrieval Examples}
The researcher conducted a qualitative analysis using several retrieval examples from the MS MARCO dataset to provide a deeper understanding of the differences between R* and baseline models. Here, comparison is done between the relevance of the top-ranked documents retrieved by R* and the baseline model.

In one example, the query was "What are the health benefits of green tea?" R* retrieved a document that directly listed the health benefits, such as antioxidant properties and improved brain function, whereas the baseline model retrieved a document that discussed green tea in general without focusing on health benefits. This demonstrates R*'s ability to prioritize documents that are more directly relevant to the specific query.

In another example, the query was "How does photosynthesis work?" R* retrieved a document that provided a step-by-step explanation of the photosynthesis process, including the light-dependent and light-independent reactions. In contrast, the baseline model retrieved a document that only briefly mentioned photosynthesis in the context of plant biology. This highlights R*'s strength in retrieving comprehensive and detailed answers.

These qualitative examples illustrate the practical improvements offered by R* in retrieving more relevant and informative documents compared to the baseline model.

\section{Conclusion}
Our study introduced R*, a novel reranking model designed to enhance document retrieval performance in retrieval-augmented generation systems. R* demonstrated superior performance on the MS MARCO dataset, underscoring the importance of model architecture and training specificity for effective reranking.

Furthermore, the comparison of R* with established models sheds light on the nuanced landscape of reranking strategies. MiniLM L6 v2's strong Recall@5 performance highlighted its ability to capture broader relevance, while the modest showing of the larger BGE model challenged the assumption that bigger models always yield better results in the context of LLMs for reranking.

Importantly, the close performance between MCQA rerankers and their cross-encoder counterparts provided empirical support for the viability of MCQA methodologies in approximating cross-encoder effectiveness for reranking. This finding underlines the significant impact that model choice and implementation can have on reranking outcomes.

Our study contributes to a deeper understanding of reranking dynamics within RAG systems, providing insights that can guide future research and development efforts. The code used in our experiments has been made publicly available to facilitate further exploration and innovation in document retrieval and reranking. By sharing these methodologies and findings, the author hopes to continue the advancement in this rapidly evolving field.

\section*{Limitations}
Our preliminary research suggests that R* tends to favor longer passages when scoring, which could introduce a bias. This is true for most cross-encoder models. It is advisable to preprocess text to normalize passage lengths for fair comparison. It is also worth noting that R* is optimized for passage-level comparisons and may not perform well on word- or phrase-level similarity tasks. The findings only apply to the MS MARCO validation data and may not generalize as well to a different dataset. Since this paper has already demonstrated a proof-of-concept, we can apply the same methodology to a larger collection of datasets for further fine-tuning. Lastly, this preliminary research is limited to open-source models and future work should include evaluation of commercially-available reranking models.

\section*{Ethics Statement}
The use of R* introduces several ethical considerations, including potential biases in the training data, privacy concerns, and the implications of automating decision-making processes. Users are encouraged to evaluate the model's fairness and transparency critically, ensuring its equitable use across diverse demographics. The author recommends that users further fine-tune this prototype model to their use case and do not use it as is, especially since this model has only been fine-tuned on MS-MARCO and not on any other domain-specific data---despite being validated on multiple datasets.

\section*{Acknowledgements}
This experimental research was written partly during Catapang's employment at Maya Philippines. However, this work is not whatsoever relevant to Catapang's non-disclosure and non-compete agreements. Furthermore, some aspects of this research was executed and would not have been possible without the support of the Tokyo University of Foreign Studies. The author extends his gratitude to his peers and the manuscript reviewers for lifting the quality of the work to the highest levels.

\bibliography{acl2023}
\bibliographystyle{acl_natbib}

\end{document}